\documentclass[aps,prl,twocolumn,superscriptaddress, nofootinbib]{revtex4-2}
\usepackage{lineno}
\usepackage{subcaption}
\usepackage{float}
\usepackage{multirow}
\usepackage{graphicx}
\usepackage{xcolor}
\usepackage{dcolumn}
\usepackage{bm}

\usepackage{bm}
\usepackage{amsmath}
\usepackage{lipsum}
\usepackage{ragged2e}
\usepackage{braket}
\usepackage{amsfonts}
\usepackage[bookmarks=false]{hyperref}
\begin{document}

\title{Designing a low-loss high-reflectivity mirror for gravitational-wave detectors by combining a dielectric metasurface and a multilayer stack}

\author{Edith Hartmann}
\email[]{edith.hartmann@fresnel.fr}
\author{Michel Lequime}
\affiliation{Aix Marseille Univ, CNRS, Centrale Med, Institut Fresnel, Marseille, France}

\author{Jérôme Degallaix}
\affiliation{Laboratoire des Matériaux Avancés—IP2I, Université de Lyon, CNRS/IN2P3, 69100 Villeurbanne, France}

\author{Michael T. Hartman}
\author{Paul Rouquette}
\author{Claude Amra}
\author{Myriam Zerrad}
\affiliation{Aix Marseille Univ, CNRS, Centrale Med, Institut Fresnel, Marseille, France}

\begin{abstract}
The design of low-mechanical-loss, high reflectivity mirrors is crucial in the development of the next generation of gravitational-wave observatories. Currently, the amorphous multilayer reflective coatings which are deposited at the surface of the test masses in interferometric gravitational-wave detectors present a major limiting factor in detector sensitivity due to their thermal noise.
These coatings require a large number of thin layers to achieve ultra-high reflectivity. However, the thermal noise generated by this type of stack increases with the number of layers used. These dielectric mirrors represent a very mature technology, with current research producing only incremental improvements, highlighting the need for new technical solutions that can address this specific issue. Here, we provide insights into the expected performance of mirrors that combine a resonant metasurface with a multilayer stack. The suggested mirror design ensures the high reflectivity required for interferometric gravitational wave detectors, while using fewer layers of properly selected materials. As a result, it significantly reduces the total coating thickness, making it a promising option for reducing thermal noise as well.

\end{abstract}

\maketitle

\section{Introduction}
\label{sec:Introduction}

Since late 2015, the scientific community has demonstrated its ability to detect gravitational waves \cite{Abbott_2016} using terrestrial observatories based on kilometer-scale Michelson interferometers equipped with extremely low-loss mirrors \cite{Harry_2010, Acernese_2015, Degallaix_2019}.
These mirrors are composed of high-quality, thick fused silica substrates coated with an all-dielectric multilayer stack that increases surface reflectivity to nearly one (typically 99.9995\%). Today, however, the thermal noise in these coatings is a limiting noise factor in the most sensitive portion (50-300 Hz) of the gravitational-wave detection band. Currently, all mirror coatings considered for interferometric gravitational-wave detectors consist of alternating layers of high-low refractive index materials, respectively titania-doped tantala (TiO$_2$:Ta$_2$O$_5$) and fused silica (SiO$_2$), deposited via ion beam sputtering (IBS).
This solution provides the best combination of refractive index contrast, low optical absorption, low thermal noise and scalability to large sizes \cite{Harry_2006}.
To improve the thermal noise performance of these coatings, a wide range of research has been conducted, a popular option being optimizing the formula of the reflective stacks to avoid using the material with the poorest thermal noise performance or minimize its thickness. For example, with mirrors that use alternated layers of silica and titania-doped tantala, the goal is to reduce the total thickness of TiO$_2$:Ta$_2$O$_5$ or replace it with higher-performing material (i.e., one with a higher refractive index, and/or lower loss angle) \cite{Principe_2015,Tait_2020,McGhee_2023,Davenport_2025}. Beyond these optimizations, several other innovative solutions are currently being investigated. These include operating at cryogenic temperatures \cite{Akutsu_2019, Craig_2019}, implementing crystalline coatings transferred onto silica substrates through optical contacting \cite{Cole_2013,Gretarsson_2025}, or proposing advanced mirror designs, such as the Khalili cavity \cite{Khalili_2005} and its etalon implementation in which the reflective function is distributed between a few coating layers on the front face of the test mass and significantly more layers on its back surface \cite{Somiya_2011}.
Maintaining the same two-sided distributed structure, Dickmann and Kroker proposed replacing the front multilayer stack with a 1D metasurface \cite{Dickmann_2018}. In our approach \cite{Hartmann_arxiv_2026, Hartmann_SPIE_2026}, the reflective structure combines a 2D metasurface and a few-layers Bragg mirror in a planar etalon configuration located entirely on the front face of the substrate.
The proposed design seeks to achieve low thermal noise while maintaining high reflectivity at $1550$ nm, thereby investigating a room-temperature alternative at this wavelength.
\begin{figure}[htpb!]
    \begin{center}
        \includegraphics[width=1\linewidth]{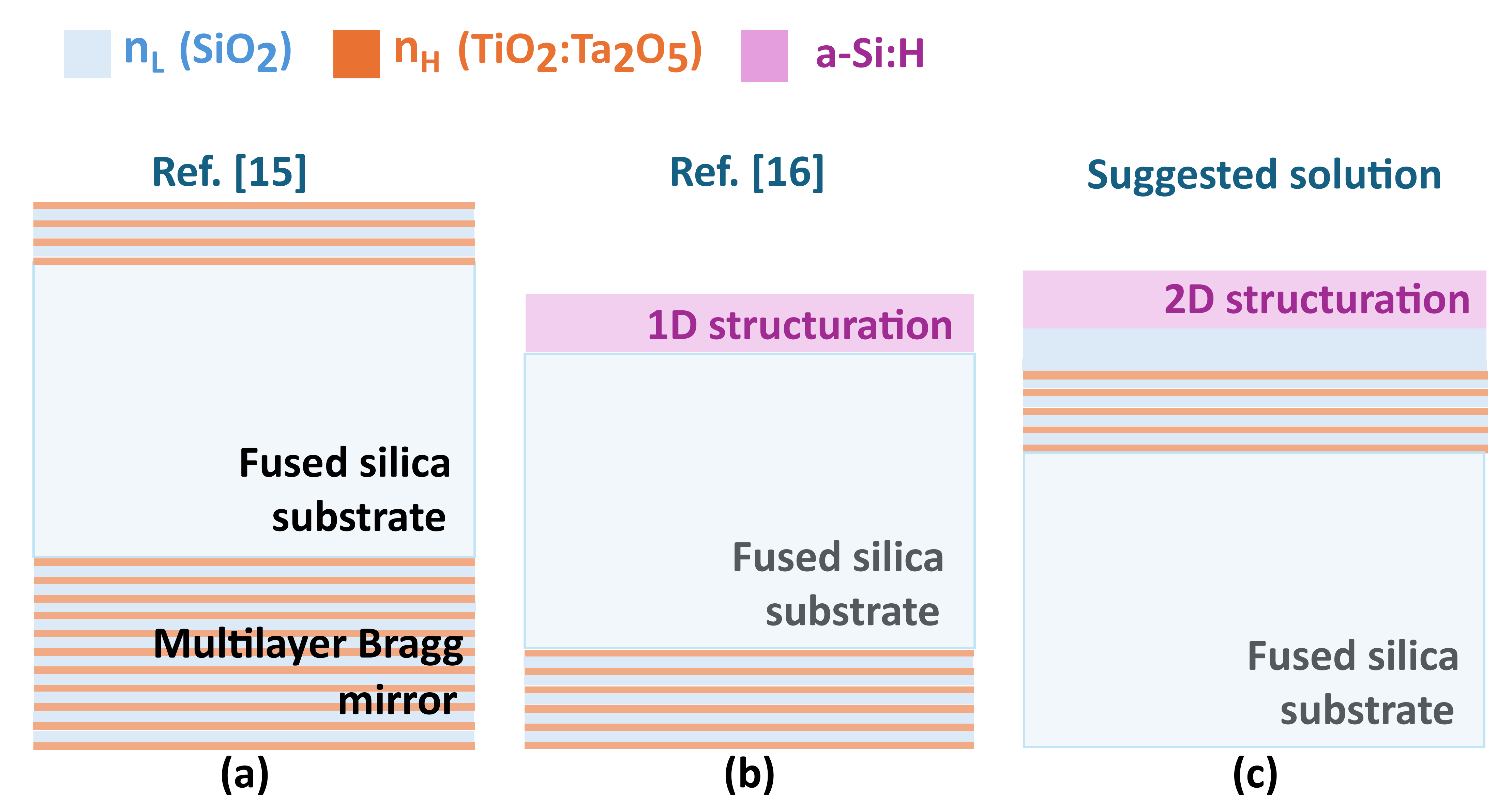}
        \caption{Schemes of advanced mirror designs: (a) Khalili etalon \cite{Somiya_2011}, (b) etalon-based meta-mirror \cite{Dickmann_2018}, and (c) our proposal.}
        \label{fig:situation}
    \end{center}
\end{figure}

Metasurfaces are planar devices structured at the wavelength scale. They have recently emerged as a new technology capable of pushing the boundaries of traditional optical components \cite{Genevet_2017, Kuznetsov_2024}, with manufacturing facilities now capable of structuring very large surfaces \cite{Zeitner_2023,Park_2024}.
Although dielectric resonant metasurfaces have theoretically demonstrated their potential as highly reflective coatings \cite{Bruckner_2010,Matiushechkina_2023}, current demonstrations are still far from what is required for gravitational-wave detectors. This is mainly because these resonant structures are highly sensitive to manufacturing errors, making every experimental realization difficult \cite{Matiushechkina_2024, Patoux_2021}. 
One way to mitigate this sensitivity is to combine the reflective metasurface with a multilayer stack, as proposed in Ref.~\cite{Dickmann_2018} and schematized on Fig.~\ref{fig:situation}b.

\begin{figure*}
\centering
\includegraphics[width=1\linewidth]{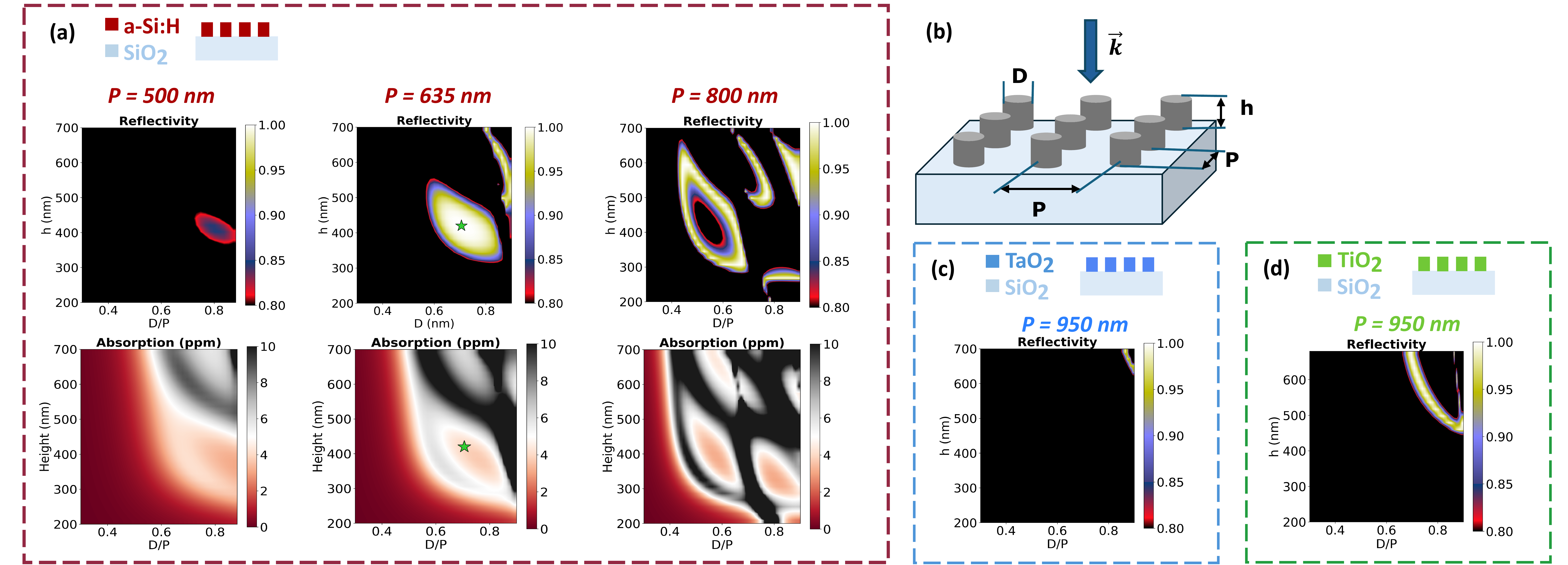}
\caption{(a) Metasurface reflectivity dependence on the periodicity $P$, the diameter $D$ and height $h$ of the pillars, when considering (a): a-Si:H, (c): tantala, or (d): titania as the metasurface material. The color code remains the same for all the cases considered here. The chosen parameters for the a-Si:H metasurface ($P= 635$ nm, $h=420$ nm and $D=448$ nm) are represented by the green star. (b) Definition of the metasurface geometric parameters; }
\label{fig:optimization}
\end{figure*}
In our approach, represented in Fig.~\ref{fig:situation}.c, a multilayer Bragg mirror is first deposited on the substrate and covered by an additional low-index layer. Next, a high-index layer is deposited on the dielectric stack and 2D-structured to form the reflective metasurface. The additional low-index layer acts as a spacer between the Bragg mirror and the reflective metasurface, with its thickness optimized to ensure anti-resonant behavior, without the need for thermal tuning.

We believe that our solution has advantages over the configuration proposed by Ref.~\cite{Dickmann_2018} in terms of practical implementation within a Michelson interferometer dedicated to the detection of gravitational waves. First, a non-negligible transmitted power through the fused silica substrate, as in Fig. 1b would promote this substrate to be part of the intracavity medium in the arm-cavities of the gravitational-wave detector. This would make the substrate manufacturing constraints much tighter in terms of rear face flatness and parallelism between the two faces. It would also be subject to the high laser fluence present in the interferometer's arms. Second, using a 1D-structured metasurface makes the mirrors sensitive to the polarization state of the light. While not critical, this adds new constraints to the already complex design and operation of the instrument. Third, the thermal tuning of the cavity required in Ref.~\cite{Dickmann_2018} must be compatible with the patterned heating of the mirror, which is already used to adjust the radius of curvature of the interferometer's end mirrors \cite{Lawrence_2004,Degallaix_2006} or stabilize the detector's operation \cite{Hardwick_2020}.

This paper therefore aims to provide insight into the design process and expected performance of the solution shown in Fig.\ref{fig:situation}.c, with a focus on reflectivity performance, manufacturing tolerances, and losses including both absorption and thermal noise.

\section{Metasurface design}
\label{sec:metasurface_design}

We present the sub-component designs of our hybrid mirror separately, starting here with the metasurface to be superposed on the mirror. It consists of polarization-insensitive nano-cylinders with identical geometries. These elements are fabricated from a high-index dielectric material and are arranged on a silica substrate ($n = 1.46$), which is assumed to be semi-infinite during the optimization.

For specific sets of geometrical parameters, these nano-cylinders can act as an array of dielectric resonators and behave as a reflective coating. In Fig.~\ref{fig:optimization}, we evaluate the reflectivity of the metasurface as a function of nano-cylinders dimensions at wavelength $\lambda = 1550$ nm, using rigorous coupled-wave analysis (RCWA) \cite{Lumerical}.
When designing such structures, a higher refractive index facilitates the achievement of this resonant behavior. Due to its high refractive index and well-established structuring processes, amorphous silicon (a-Si) is one of the most promising candidates for the metasurface material. However, absorption in a-Si is not negligible and needs to be carefully evaluated, especially for applications such as gravitational-wave detection, which requires incredibly low absorption levels (typically below the ppm). The extinction coefficient of amorphous silicon depends largely on the deposition process and varies widely in the recent literature, ranging from $10^{-3}$ to $10^{-5}$ for the best realizations reported to our knowledge \cite{Steinlechner_2016,Birney_2018}. However, in the early 2000s, Domash et al. obtained much better results with hydrogenated amorphous silicon (a-Si:H) deposited by Plasma Enhanced Chemical Vapor Deposition (PECVD) in the context of designing narrow-band filters with thermally tunable central wavelength and a-Si:H as the spacer material \cite{Domash_2004}. Experimental measurements confirmed extremely low absorption, with an extinction coefficient as low as $10^{-6}$ at wavelength 1500 nm \cite{Domash_2003}. Additionally, recent studies seem to confirm the feasibility of achieving such a low extinction coefficient for a-Si:H using PECVD as the deposition process \cite{Tawalare_2021, Takei_2014}. Hydrogenation of a-Si deposited by other methods also gives promising results and is currently under investigation in the context of gravitational-wave detector \cite{Molina-Ruiz_2023, Molina-Ruiz_2025}.

Considering a-Si:H ($n = 3.67$, $k=10^{-6}$ \cite{Domash_2003}) as the pillar material, the period $P$ of the metasurface is optimized to ensure high reflectivity over the broadest possible parameter range, thereby enhancing fabrication tolerances. Figure \ref{fig:optimization}a illustrates how the period influences the region exhibiting high reflectivity: as the period increases, this region expands before eventually degenerating into a ring. Therefore, we select a period of $P=635$ nm, which corresponds to the widest region prior to degeneration.
The pillars diameter $D$ and height $h$ are then chosen to lie at the center of this high-reflectivity region, respectively $h=420$ nm and $D=448$ nm as represented by the green star in Fig.~\ref{fig:optimization}a.

It is no longer possible to obtain this kind of behavior (growth of a stability region before degeneration) when the refractive index of the material used to manufacture the metasurface is lower. This is illustrated in Fig.~\ref{fig:optimization}c with tantala ($n = 2.07$) and in Fig.~\ref{fig:optimization}d for titania ($n = 2.33$) \cite{Granata_2020}.

\begin{figure}[htbp]
    \centering
    \includegraphics[width=1\linewidth]{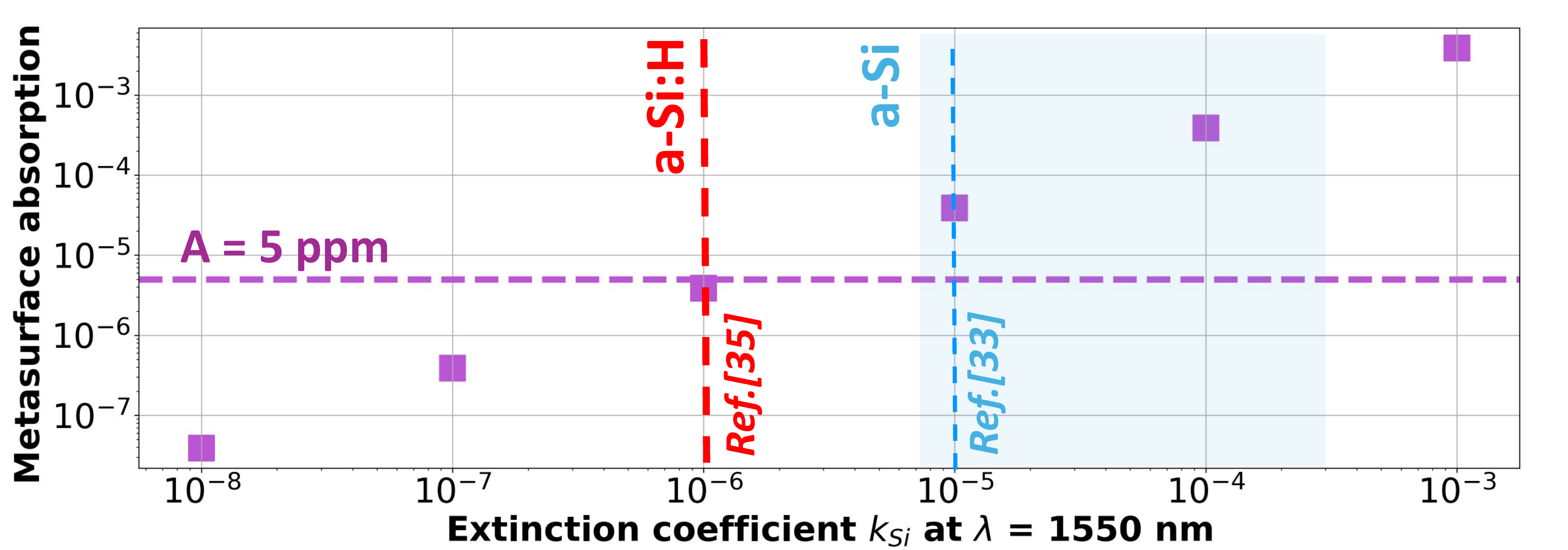}
    \caption{Influence of the a-Si extinction coefficient on the absorption of the optimized metasurface (green star on Fig.~\ref{fig:optimization}a ).}
    \label{fig:metasurface_sensibility}
\end{figure}

The absorption of the a-Si:H metasurface, calculated as $A=1-R-T$, is also closely related to the resonant behavior, exhibiting the same overall trends as the reflectivity (see Fig.~\ref{fig:optimization}a). For the optimized design, the absorption is predicted to be approximately 4 ppm while maintaining a reflectance exceeding 99.999$\%$.
Since achieving an extinction coefficient of $10^{-6}$ remains technologically challenging, Fig.\ref{fig:metasurface_sensibility} shows the predicted absorption as a function of the extinction coefficient. These results highlight the importance of precisely controlling the deposition process of the a-Si:H layer in order to minimize optical absorption in the material.

It should be noted that a-Si:H could not be considered for operation at 1064 nm, owing to its high optical absorption at this wavelength. While it would be highly desirable to develop a similar hybrid design for 1064 nm (the operating wavelength of current gravitational-wave detector), we are not aware of any material that offers a sufficiently high refractive index and an adequately low extinction coefficient at this wavelength. Titania at 1064 nm could be an interesting solution (see Supplementary Materials), but the achievable reflectivity performance remains well below that of aSi:H at 1550 nm and its intrinsic thermal noise performance is poor (loss angle greater than $10^{-3}$) \cite{Granata_2020}.
Consequently, in the following we focus on the a-Si:H metasurface designed for operation at 1550 nm

\begin{figure*}
    \centering
    \includegraphics[width=1\linewidth]{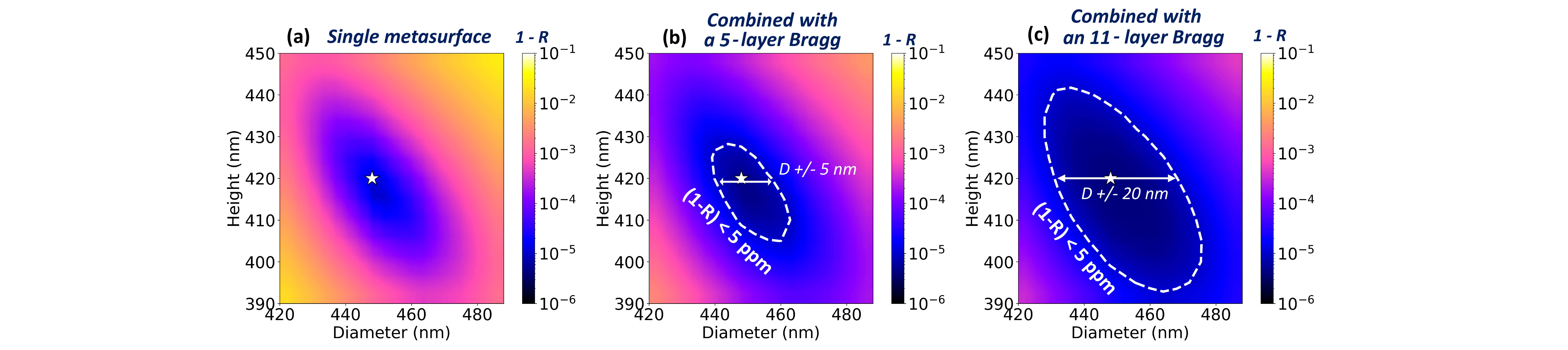}
    \caption{Manufacturing tolerances: $1-R$ as a function of the metasurface geometric parameters for either the single metasurface (a), or when combined with a TiO$_2$:Ta$_2$O$_5$/SiO$_2$ Bragg mirror of 5 (b) and 11 (c) layers. The white dashed lines show the area where $1-R<5$ ppm. }
    \label{fig:tolerances}
\end{figure*}

\section{Association with a conventional multilayer Bragg mirror}
\label{sec:AssociationwithaConventionalMultilayerBraggMirror}

Because of the resonant nature of the metasurface, its reflectivity is highly sensitive to variations in the pillar geometry. Even slight deviations in the pillar dimensions can induce a significant reduction in reflectivity, making it challenging to ensure the required reflectivity performance ($R> 99.999 \%$) across the full range of typical fabrication tolerances. The robustness of these structures can however be improved by coupling the metasurface with a multilayer stack.

In our approach illustrated in Fig.~\ref{fig:situation}b, the anti-resonant behavior of the hybrid device is achieved by optimizing the thickness $e$ of the additional low-index layer. This coupling layer acts as the spacer of the Fabry–Perot cavity formed between the two reflective structures, namely the metasurface and the multilayer mirror.
For simplicity, we consider the multilayer to be an ideal Bragg mirror, with quarter-wave layers of alternating high and low index materials, such as TiO$_2$:Ta$_2$O$_5$ ($n = 2.08$, corresponding to a Ti/Ta mixing ratio of $27\%$ \cite{Granata_2020}) and SiO$_2$. The reflection phase shift on such a quarter-wave multilayer stack is equal to $\pi$ at the design wavelength, while that on the optimized metasurface is equal to $1.066\pi$, making the anti-resonant condition achieved at $\lambda = 1550$ nm  with a silica spacer thickness of $e=779$ nm. Further details on the determination of the optimal silica spacer thickness between the Bragg mirror and the metasurface are provided in the Supplementary Materials. \\

Finally, the number of layers in the Bragg mirror is determined by the expected fabrication tolerances of the metasurface.
Figure \ref{fig:tolerances} shows the performance of the reflective structure when slight changes are made to the geometrical parameters of the metasurface, $D$ and $h$. This analysis is shown either for a single a-Si:H metasurface, and when coupled with a Bragg mirror of 5 and 11 TiO$_2$:Ta$_2$O$_5$/SiO$_2$ layers (in total). The white dashed lines in Fig.~\ref{fig:tolerances} show the area where the losses $(1-R)$ are below 5 ppm. Although some parameters locally fulfill these conditions for the single metasurface case (Fig.~\ref{fig:tolerances}a), this area is extremely small and is not represented for the sake of readability. However, as the number of layers used in the Bragg mirror increases, this dotted area expands, highlighting how coupling with the multilayer relaxes the metasurface manufacturing tolerances.
Our design strategy is therefore to select the number of layers in the Bragg mirror according to the anticipated fabrication tolerances. For example, 5-layer and 11-layer Bragg mirrors respectively allow to compensate for target pillar diameter deviations of approximately $\pm5$ nm and $\pm20$ nm,  making it much more realistic to achieve the required performance in practice.

\section{Discussion on losses}
\label{sec: losses}
As mentioned previously, the aSi:H metasurface exhibits an absorption of approximately 4 ppm with $k=10^{-6}$. Moreover, because most of the incident light is reflected by the metasurface itself, only a small fraction propagates into the Bragg mirror. As a result, the change in the layers number of the Bragg mirror has no impact on the absorption performance of the hybrid design, which is essentially equal to that of the metasurface alone.
Note that the absorption level obtained here is still high compared to current standards for gravitational-wave detectors. Improving the deposition process of a-Si:H, or exploring alternative high-index materials, could help achieve the target absorption levels.

Nevertheless, we believe that the potential of such hybrid approaches to reduce coating thickness makes them a promising direction for developing ultra low-noise mirror. In the present design, the multilayer thickness is reduced by approximately a factor of three compared with a conventional Bragg mirror, while requiring only 11 layers (5 pairs). This reduction is particularly attractive because the Brownian thermal noise of a thin-film coating scales with its thickness \cite{Principe_2015}.
However, unlike in a conventional Bragg mirror, here most of the incident field is reflected by the metasurface itself, making its thermal-noise contribution potentially more significant than that of the underlying multilayer. The thermal noise of such a nanostructured layer is less straightforward to evaluate and requires coupled electromagnetic and mechanical numerical simulations \cite{Kroker_2017, Dickmann_2018_2}. Ref.~\cite{Gaedtke_2026} highlights the potential of metasurfaces and compares the thermal noise performance of different meta-atom geometries. It concludes that nanocylinder-based metasurfaces, such as those considered here, can achieve thermal noise levels comparable to those of ridge-based designs.
These results therefore suggest that our design should exhibit thermal noise performance similar to that predicted for hybrid mirrors incorporating ridge-based metasurfaces, as investigated in Refs. \cite{Dickmann_2018,Kranhold_2026}, the latter of which was brought to our attention only recently. Nevertheless, a comprehensive thermal noise analysis of our design, and especially of the a-Si:H metasurface is still underway. More generally, it should be noted that all current thermal noise predictions and numerical simulations reported so far for metasurface-based mirrors still require experimental validation based on a direct measurement of the thermal noise \cite{Gras_2018}, which has not yet been demonstrated.

Finally, uncontrolled scattering caused by surface roughness may also constitute a source of optical loss. While the surface roughness of the Bragg mirror layers deposited using well-established processes such as IBS is well known and tightly controlled (perfect replication of the roughness of the substrate), that of the metasurface may represent a more significant source of loss, as suggested in \cite{Dickmann_2023}. Indeed, large-scale surface roughness, with spatial frequencies within the optical range, may induce measurable scattering losses \cite{Lequime_2018,Fouchier_2020}. More importantly, much smaller-scale surface roughness and fabrication-induced irregularities of the meta-atoms may also degrade the reflective performance of the metasurface, thereby introducing additional losses \cite{Kranhold_2026}. We are currently developing an electromagnetic model based on the differential method \cite{Arnaud_2008} to quantify all these effects, with the aim of comparing its predictions with measurements.

\section{Conclusion}
\label{sec:Conclusion}

In conclusion, we investigated the feasibility of combining a resonant a-Si:H metasurface and a multilayer Bragg mirror. These two reflective structures are coupled through an additional silica layer, whose thickness is optimized to ensure maximum reflectivity. This hybrid approach significantly improves the robustness of the metasurface reflectivity to fabrication errors, paving the way toward more accessible experimental realization.
The proposed mirror solution can theoretically achieve the required performance for gravitational-wave detection applications, while significantly reducing the coating thickness, making it a promising solution to potentially reduce the thermal noise as well. The absorption of such resonant metasurface needs to be carefully monitored, as well as potential scattering losses.

Although experimental validation is still required, we strongly believe that such hybrid designs are promising enough to encourage further investigations. Beyond representing a major challenge for the next generation of gravitational-wave detectors, the development of mirror coatings with reduced thermal noise also benefits other high-precision measurement experiments that rely on ultra-stable laser cavities. Examples include the realization of highly accurate optical atomic clocks \cite{Jiang_2011,Wislo_2017} or the investigation of vacuum magnetic birefringence \cite{Agil_2022,Zavattini_2018}.

\begin{acknowledgments}
\textit{Acknowledgments} This work has been achieved thanks to the support of the French Centre National d’Etudes Spatiales - CNES and the company CILAS in the context of the joint laboratory LabTOP, the Virgo collaboration, A*MIDEX foundation for the funding of GRAVITERM project (AMX-22-RE-AB-091), and Agence Nationale de la Recherche for the funding of X-LOSM project (ANR-23-CE08-0027).
\end{acknowledgments}

\newpage
\begin{appendix}

    \section{Appendix 1:  Coupling Layer Thickness}
\begin{figure}[htpb]
    \centering
    \includegraphics[width=1\linewidth]{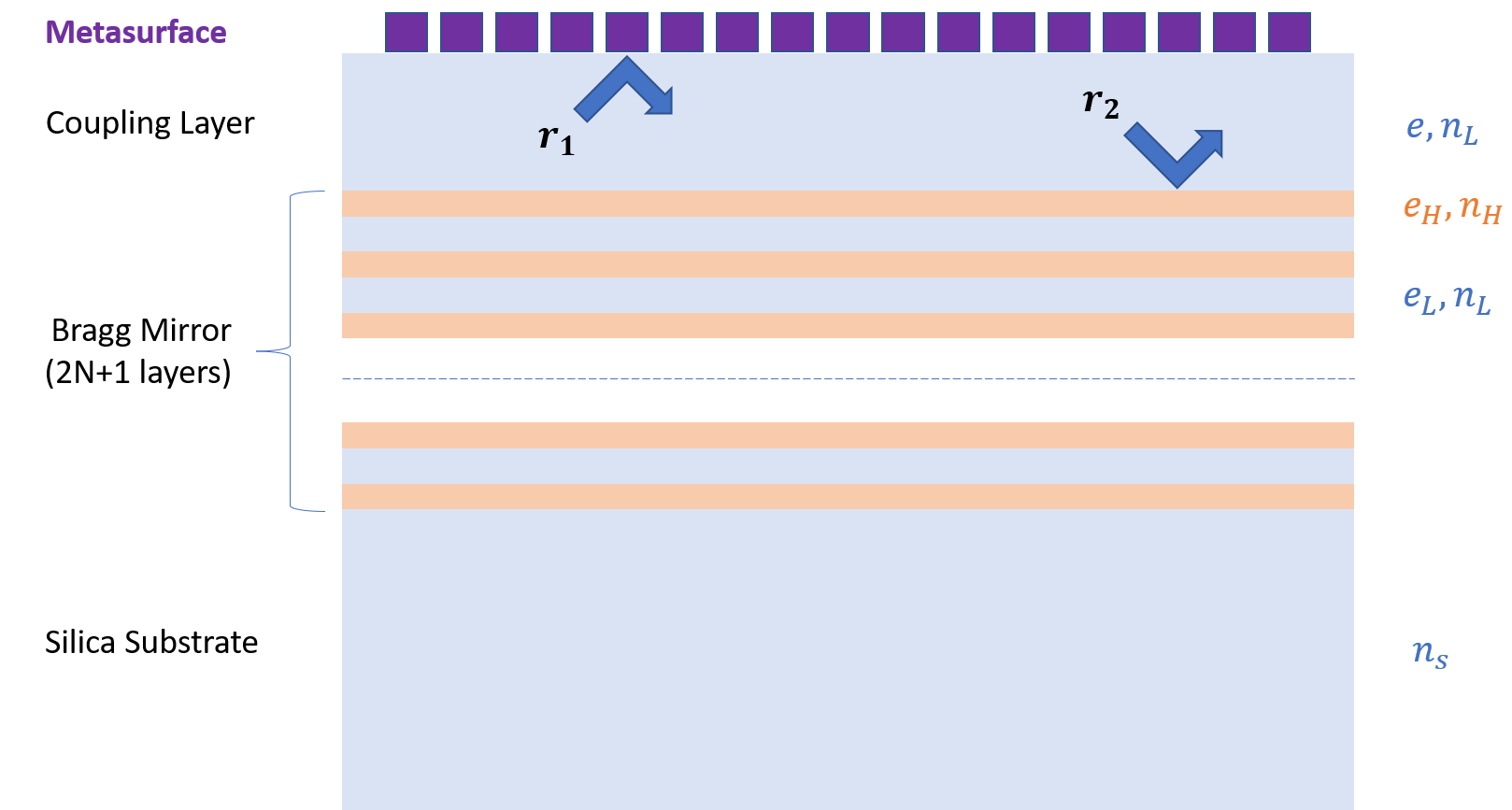}
    \caption{Schematic representation of the hybrid mirror}
    \label{fig:Hybrid Mirror}
\end{figure}

We consider the hybrid mirror design illustrated in Fig.\ref{fig:Hybrid Mirror}, which consists of two reflective structures (a metasurface with a reflection coefficient $r_1$ and a multilayer stack with a reflection coefficient $r_2$), separated by a coupling silica layer of thickness $e$. This structure can be considered as a planar Fabry-Perot, whose transmittance $T_{\text{FP}}$ is defined as follows:
\begin{align}
&T_{\text{FP}}=\frac{(1-R_1)(1-R_2)}{1+R_1R_2-2\sqrt{R_1R_2}\cos\Phi}\quad\\
&\text{with}\quad r_1=\sqrt{R_1}\thinspace e^{i\rho_1}\quad\text{and}\quad r_2=\sqrt{R_2}\thinspace e^{i\rho_2}
\end{align}
where $\Phi$ is the round-trip phase of the cavity:
\begin{equation}
\Phi=\frac{4\pi e n_{\text{SiO}_2}}{\lambda}+\rho_1+\rho_2
\end{equation}

For simplicity, we will consider the multilayer to be an ideal Bragg mirror, with quarter-wave layers of alternating high- and low-index materials, such as TiO$_2$:Ta$_2$O$_5$ ($n_H = 2.08$ at $\lambda = 1550$ nm, corresponding to a Ti/Ta mixing ratio of $27\%$ \cite{Granata_2020}) and SiO$_2$ ($n_L=1.46$ at $\lambda=1550$ nm). For such a quarter-wave multilayer stack, we have \cite{Amra_2021}
\begin{equation}
r_2=\frac{n_L-Y_{BM}}{n_L+Y_{BM}}
\end{equation}
where $Y_{BM}$ is the complex admittance of the top interface of the Bragg mirror, defined at the design wavelength and at zero angle of incidence by:
\begin{equation}
Y_{BM}=\frac{n_H^2}{n_s}\left[\frac{n_H}{n_L}\right]^{2N}>n_H\quad\forall N\quad (n_s=1.46)
\end{equation}
As a consequence, $r_2$ is real, but negative, leading to $\rho_2 = \pi$.

We will now assume that the phase change $\rho_1$ at the reflection on the metasurface is weakly dependent on the thickness $e$, and use, as a first approximation, a constant value that corresponds to the case where this thickness is infinite (we will verify this assumption further). This constant value, denoted $\rho_1(\infty)$, is determined by RCWA simulation and is equal to $1.066\pi$.

To ensure an anti-resonant behavior to the Fabry-Perot (i.e. a maximum reflectivity for the hybrid mirror), the round-trip phase must be an odd integer multiple of $\pi$. This can be achieved with an infinite number of thicknesses, $e_k$, that satisfy the following relation: 
\begin{align}
e_k&=\frac{\lambda}{4\pi n_{\text{SiO}_2}}\{(2k+1)\pi-[\rho_1(\infty)+\rho_2]\}\\
&=\frac{\lambda}{4n_{\text{SiO2}}}\left[2k-\frac{\rho_1(\infty)}{\pi}\right]\quad k\in \mathbb{N}
\label{eq:cavity_thickness}
\end{align}
These thicknesses are spaced by an interval of $\lambda/(2n_{\text{SiO}_2})$ and depend only on the phase term $\rho_1(\infty)$ acquired at reflection on the metasurface. For example, we have: $k=1,\thinspace e_1=247.9$ nm, $k=2,\thinspace e_2=778.8$ nm, and so on.

Consequently, the optimal thickness solutions $e_k$ are unaffected by the number of layers used for the Bragg mirror. However, this conclusion is based on the assumption that the phase change at reflection on the metasurface is independent of the coupling layer thickness $e$. To validate this hypothesis, we studied the influence of this thickness on the hybrid mirror transmittance through RCWA simulations of the entire assembly.

\begin{figure}[htpb]
    \centering
    \includegraphics[width=1\linewidth]{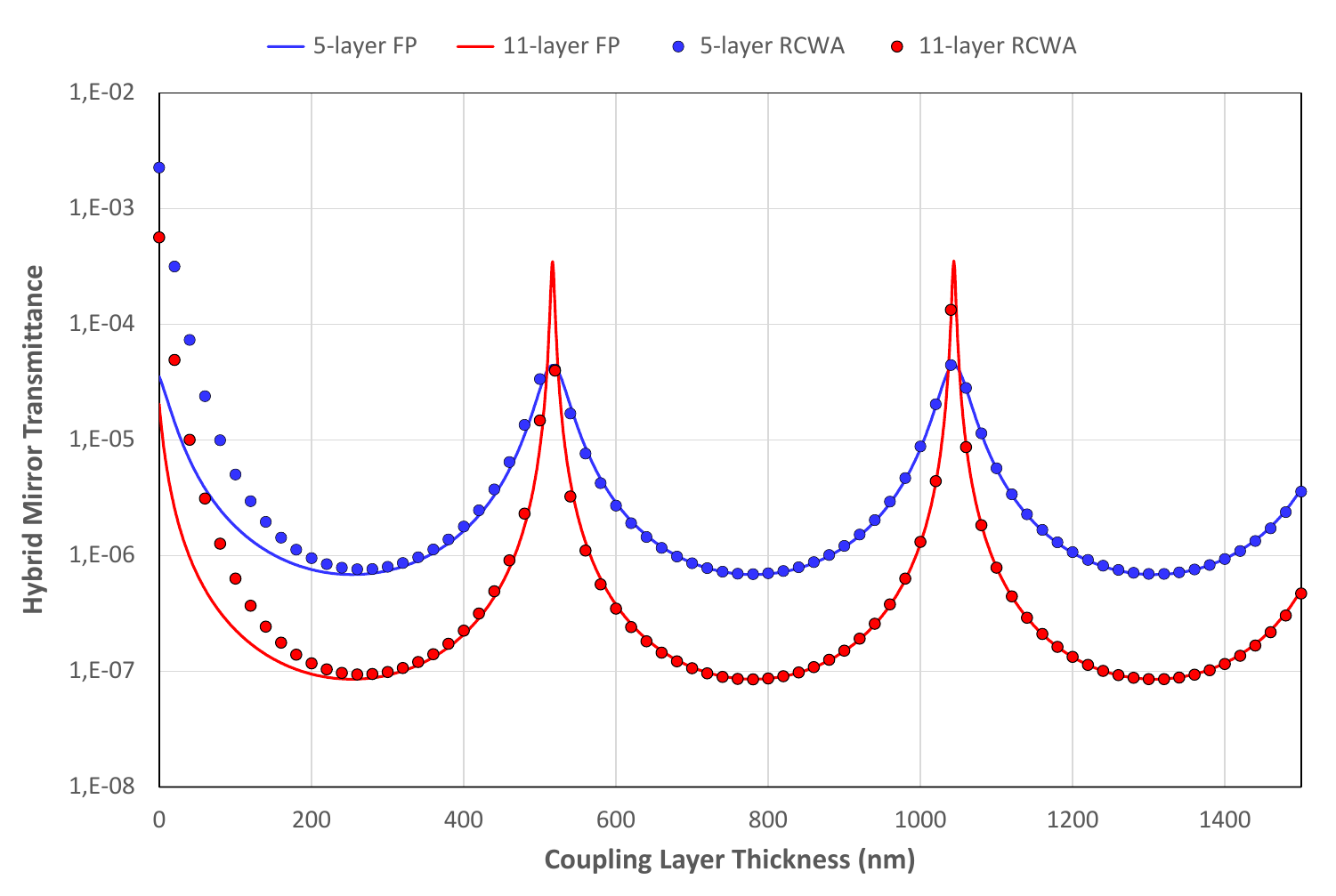}
    \caption{Hybrid mirror transmittance as a function of the silica spacer thickness $e$. The figure shows the simulated values (colored dots) and the values predicted by the Fabry-Perot model (solid lines), for mirror designs combining the optimized a-Si:H metasurface with a Bragg mirror of either 5 or 11 layers. The selected thickness is $e=778.8$ nm.}
    \label{fig:cavity_thickness}
\end{figure}
The simulated transmittance values are shown in Fig. \ref{fig:cavity_thickness} in comparison with those resulting from the Fabry-Perot model using
\begin{equation}
R_1=0.999994\quad\text{;}\quad R_2=|r_2|^2
\end{equation}
i.e. $R_2=0.618$ for the 5-layer Bragg mirror and $R_2=0.944$ for the 11-layer Bragg mirror. All the simulated values are identical to those provided by the Fabry-Perot model, except for coupling layer thicknesses smaller than 300 nm. We believe this is mainly because the metasurface reflection coefficient $r_1$ is obtained with the structured layer placed on a semi-infinite SiO$_2$ substrate. In the assembly, however, the metasurface is placed on a silica layer, which causes the Fabry-Perot model to be less accurate. As the coupling layer thickness increases, the semi-infinite substrate hypothesis used to optimize the metasurface quickly becomes valid, and the assembly response converges to the Fabry-Perot model. 

Since smaller cavity thicknesses are preferable in terms of thermal noise, it could be interesting to select the first solution of equation (\ref{eq:cavity_thickness}), i.e. $e=247.9$ nm. However, with this choice, the Fabry-Perot model is not fully accurate. Therefore, it seems preferable to consider the second solution of equation (\ref{eq:cavity_thickness}), namely $e=778.8$ nm for which the RCWA simulation and the Fabry-Perot model lead to the same result. Additionally, the first thickness solution does not significantly reduce thermal noise, because silica exhibits excellent mechanical properties and contributes minimally to the overall coating thermal noise.

Finally, because the selected thickness is at the center of the stable plateaus predicted by the Fabry–Perot model, which allows for greater stability in the event of potential thickness errors when depositing the coupling silica layer.

\section{Appendix 2: Design at 1064 nm}

Current and most planned future gravitational-wave detectors operate at a wavelength of $\lambda = 1064$ nm, making it desirable to develop the same type of hybrid design for this wavelength.
However, a-Si:H is not a suitable material for operation at 1064 nm because of its high optical absorption. Titanium dioxide (TiO$_2$) instead appear as a promising candidate for the metasurface material owing to its relatively high refractive index ($n= 2.35$ at $\lambda=1064$ nm \cite{Granata_2020}) and negligible optical absorption (here we consider $k = 10^{-7}$).

\begin{figure*}
    \centering
    \includegraphics[width=0.7\linewidth]{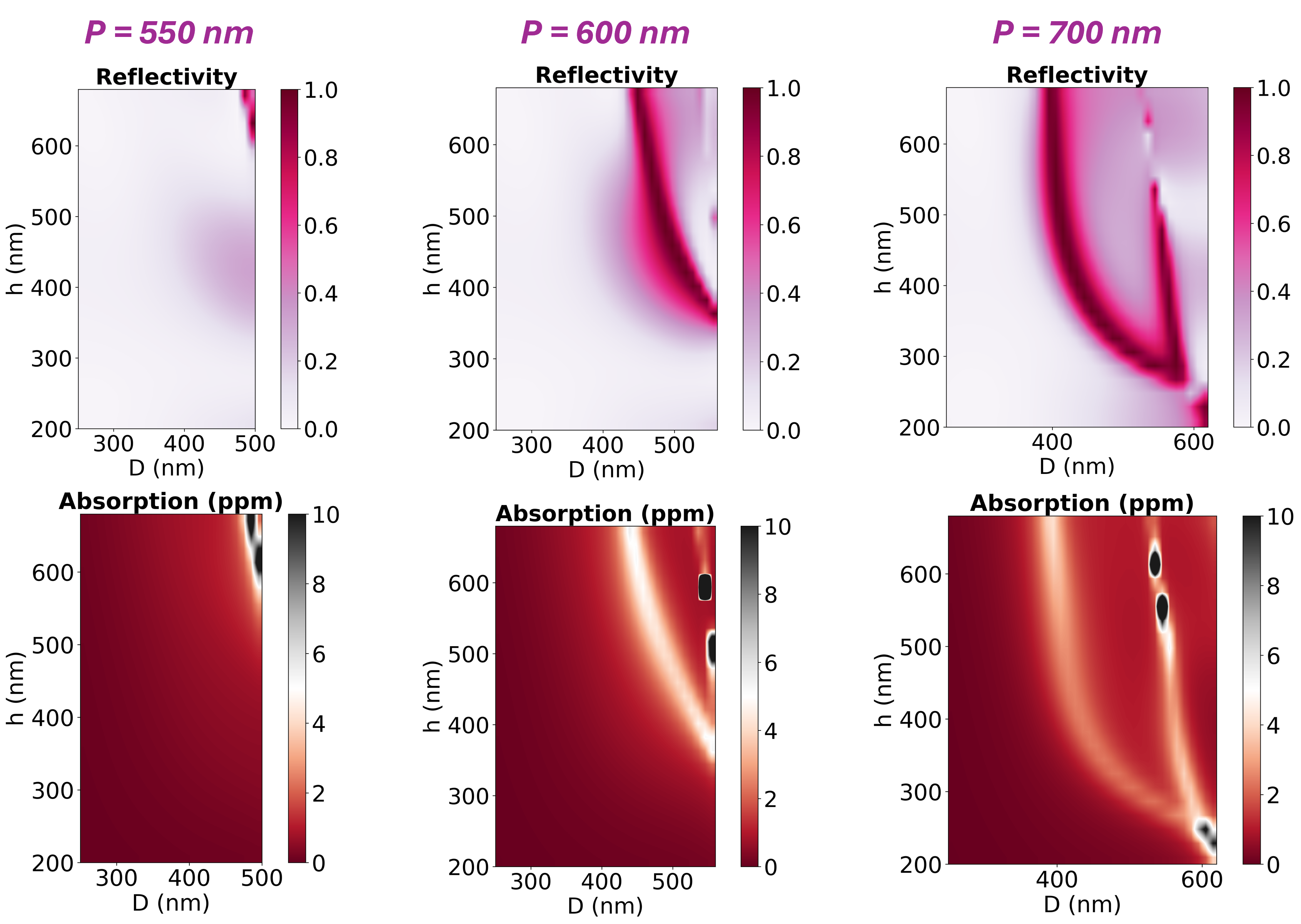}
    \caption{Metasurface performance at $\lambda = 1064$ nm as a function of the periodicity $P$, the pillars diameter $D$ and height $h$, when considering titania as the metasurface material.}
    \label{fig:1064}
\end{figure*}
Fig. \ref{fig:1064} presents the optimization results for a TiO$_2$ metasurface deposited on a SiO$_2$ substrate, at wavelength 1064 nm. Suitable geometrical parameters ($h$, $D$, $P$) can be identified that provide high reflectivity. However, unlike the a-Si:H metasurface designed for 1550 nm, it was not possible to reduce the period $P$ sufficiently to obtain a broad and robust high-reflectivity region. As a result, the TiO$_2$ design exhibits a more limited tolerance to fabrication imperfections.
Furthermore, the intrinsic thermal noise performance of TiO$_2$ remains relatively poor (loss angle greater than $10^{-3}$ \cite{Granata_2020}. 

Consequently, in this manuscript we focus on the a-Si:H metasurface designed for operation at 1550 nm.

\end{appendix}

\end{document}